\documentclass[twocolumn,superscriptaddress,showpacs,prl]{revtex4}

\usepackage{amsmath,amssymb,latexsym,graphics,epsfig}
\usepackage{pstricks}
\usepackage{float}

\newpsobject{showgrid}{psgrid}{subgriddiv=1}

\begin{document}

\title{Probability of the emergence of helical precipitation patterns in the
    wake of reaction-diffusion fronts}

\author{Shibi Thomas}
\altaffiliation[Present address: ] {Department of Physics, University of Calicut, 673635 Kerala, India}
\email{shibithomas969@gmail.com}
\affiliation{Department of Theoretical Physics,  E\"{o}tv\"{o}s University, 1117 Budapest, Hungary}

\author{Istv\'an Lagzi}
\email{istvanlagzi@gmail.com}
\affiliation{Department of Physics, Budapest University of Technology and Economics, 1111 Budapest, Hungary}

\author{Ferenc Moln\'ar Jr.}
\email{molnaf@rpi.edu}
\affiliation{Department of Physics, Applied Physics, and Astronomy, Rensselaer Polytechnic Institute, 12180 Troy, NY, USA}

\author{Zolt\'an R\'acz}
\email{racz@general.elte.hu}
\affiliation{Institute for Theoretical Physics - HAS,
  E\"otv\"os University, 1117 Budapest, Hungary}

\date{\today}

\begin{abstract}
Helical and helicoidal precipitation patterns emerging in the 
wake of reaction-diffusion fronts are studied. 
In our experiments, these chiral structures arise
with well defined probabilities $P_{\rm H}$ controlled by 
conditions such as e.g., the initial concentration of the reagents. 
We develop a model which describes the observed experimental trends. 
The results suggest that $P_{\rm H}$ is determined by a 
delicate interplay among the time- and length-scales related 
to the front and to the unstable precipitation modes 
and, furthermore, 
the noise amplitude also plays a quantifiable role. 
\end{abstract}
\pacs{05.40.-a, 02.50.-r, 68.35.Ct}

\maketitle


Helices and helicoids are present from nano- to macro-scale (ZnO 
nanohelices \cite{Gao-Zincoxid}, macromolecules and inorganic crystals 
with a helical structure \cite{Imai-chiral,Su-doublehelix}, precipitation helices \cite{Sci-1982-Muller,helical-ribbons,Giraldo2000}, fiber 
geometry of heart walls \cite{heart-helix}). Formation of these
fascinating structures generally follows two routes. First, 
templates with chiral symmetry (e.g., oragogel fibers) may exist 
in the system, and the symmetry is just transcribed to a 
structure (e.g., inorganic materials \cite{chiral-fibers})
at a larger scale. Second, spontaneous symmetry breaking 
may occur through the self-assembly of achiral building blocks 
into a helical/helicoidal structure, as e.g. in case of crystals 
with chiral morphology \cite{Imai-chiral,chiral-blocks}. 

Theoretically, the symmetry-breaking route is 
more interesting. Universal aspects may emerge and 
the robust features of this 
self-organization process may be important for 
applications as well. Indeed, control over creating helical 
structures would make engineering (in particular, the 
bottom-up design of micro-patterns \cite{Grzy2005}) more flexible 
since chiral morphology of materials are known to affect their 
physical (electronic) properties \cite{Giraldo2000,Carbontube}.

In order to develop insight into the genesis of helices/helicoids, 
we investigate an emblematic example of pattern formation, 
namely the formation of precipitation patterns in the wake of reaction-diffusion fronts \cite{Henisch,MullerRoss2003}. 
The motivation for this choice comes from the observation
that helicoidal structures have an axis, and the correlations are simple
in the plane perpendicular to the axis. 
This suggests that building the perpendicular correlations 
in the wake of an advancing planar front may be  
a simple and natural mechanism of creating helices/helicoids.
Additional motivation comes from the existence of a large body
of knowledge in the related Liesegang phenomena 
\cite{Henisch,MullerRoss2003}. It allows 
the use of well-established experimental and
theoretical approaches, thus making it easier to develop
a novel view on the formation of 
helical structures.

Our main results concern the probabilistic
aspects of the symmetry-breaking route. We determine  
the probability $P_{\rm H}$ of the emergence of single
helices/helicoids in Liesegang-type experiments as the conditions 
such as the initial concentration of inner or outer 
electrolytes, or the temperature are changed. 
$P_{\rm H}$ is found to be well reproducible and
large ($P_{\rm H}>0.5$ for some parameter range). 
The results are understood by expanding and simulating 
a model of formation of precipitation patterns \cite{ModelB1999}.
We explicitly observe that the origin of 
helices/helicoids is not to be found in the fluctuations and 
asymmetry of the initial- or boundary conditions \cite{Polezhaev1991,Polezhaev1994}. Instead, the growth of 
unstable modes, the dynamics of the front, 
and the bulk fluctuations (noise) combine to yield the
helices.

\begin{figure}[htb]
  \includegraphics[width=\columnwidth]{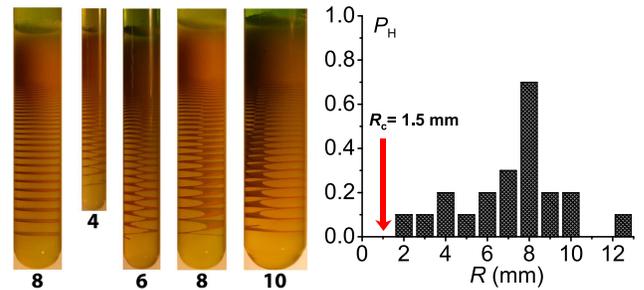}
\caption{Regular Liesegang- (leftmost tube) and helicoidal 
patterns (all other tubes) in agarose gel with the numbers 
corresponding to the tube radius $R$ measured in mm. 
$R$ is varied at fixed experimental 
conditions ($T$=22$\, ^{\rm o}$C, [Cu$^{2+}$]$_0$=$a_0$=0.5 M, and 
[CrO$_4^{2-}$]$_0$=$b_0$=0.01 M) and the probability of 
helicoid formation $P_{\rm H}$ is displayed (right 
panel). No helicoid appears for $R\leq R_c$.
\label{F:helix-exp}}
\end{figure} 
In our experiments, we study the precipitation reaction 
$\rm Cu^{2+}(aq)+CrO_4^{2-}(aq)\to CuCrO_4(s)$ in 1\% agarose gel. 
The gel soaked with $\rm K_2CrO_4$ (inner electrolyte) is placed 
in a test tube and a solution of $\rm CuCl_2$ (outer electrolyte) 
is poured on top of the gel. Setting the concentration of the outer 
electrolyte an order of magnitude larger than that of the inner 
electrolyte yields a reaction front diffusing into the gel, and 
a Liesegang pattern of precipitation bands forms behind the front
(Fig.\ref{F:helix-exp}). Frequently, however, helicoids 
evolve from the same macroscopic experimental conditions (Fig.\ref{F:helix-exp}). 
We quantified the stochastic nature of this 
intriguing phenomenon by varying the 
concentration of the outer ($a_0$) and inner ($b_0$) electrolytes, the 
radius of the test tube ($R$), and the temperature ($T$), 
and measuring $P_{\rm H}$ 
using 10 independent experiments for each parameter set. 

\begin{figure}[htb]
  \includegraphics[width=\columnwidth]{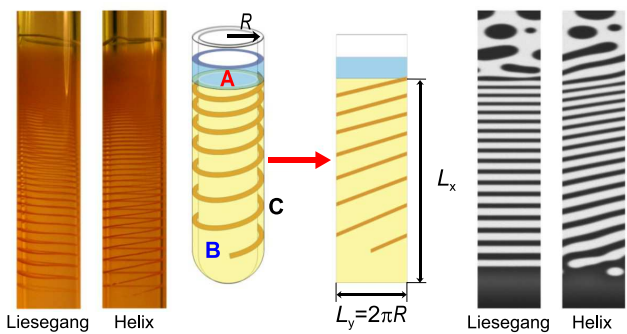}
\caption{Transforming the thin layer of gel in the {\it 
tube-in-tube} experiment into a two-dimensional strip. 
The Liesegang bands and the helices in the experiments were 
obtained in agarose gel at $T = 22\, ^{\rm o}$C, $a_0 = 0.5$ M, 
and $b_0$ = 0.01 M, with the radii of the outer and inner tubes 
being 8 and 6 mm. The scaled parameters used for the simulations 
(columns on the right with the precipitate shown in white) 
were $a_0=80$, $b_0=1$, 
$\sigma =0.8$, $\lambda =0.2$, $\eta =0.05$, and $L_y=64$. 
\label{F:helix-plain}}
\end{figure}

Similar experiments were carried out in a quasi two-dimensional 
geometry as well. The gel (with the inner electrolyte $B$) was 
placed in the gap between two test tubes of slightly different 
radii ($\delta R =2$ mm), thus effectively confining the pattern 
to the surface of a cylinder (Fig.\ref{F:helix-plain}). 
In this geometry, we observed the formation of 
regular Liesegang rings, single helices, double helices, 
and more complex patterns for large $R$.

Fig.\ref{F:helix-exp} shows regular bands and helicoidal 
patterns in test tubes of various radii, together with the measured 
probability of helicoid formation ($P_{\rm H}$). 
We observe no helicoids below a critical radius ($R_c$=1.5 mm), 
in agreement with theoretical expectations based on a simplified 
model where the reaction front moves with fixed velocity \cite{Polezhaev1991,Polezhaev1994}. For $R>R_c$, one finds 
that $P_{\rm H}$ increases with increasing $R$, and 
reaches rather large values ($P_{\rm H} \approx 0.7$ at $R= 8$ mm) 
before decreasing again. 
The decrease at large $R$ is due to the noticeable 
proliferation of complex structures (double or triple 
helicoids, disordered patterns) which suppress the weight 
of single helicoids. 

\begin{figure*}[ht!]
{\includegraphics[width=\textwidth]{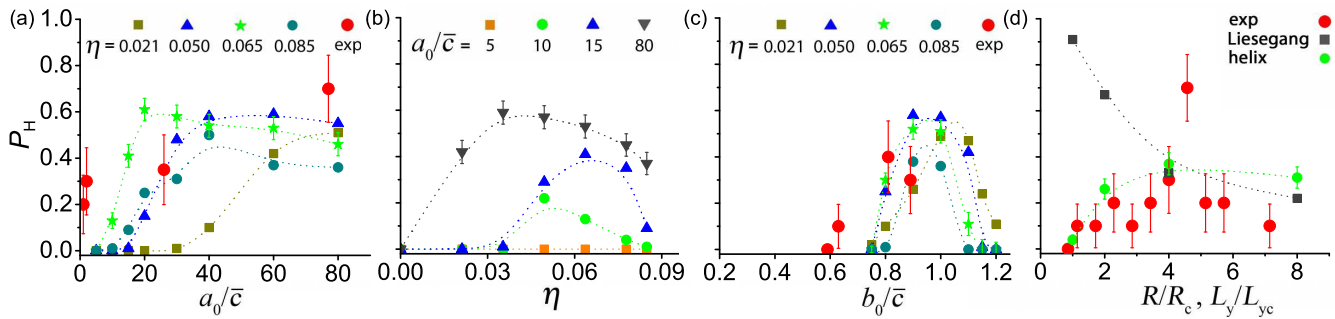}}
\caption{The probability of helicoid/helix formation 
$P_{\rm H}$ in experiments (large red dots) and in simulations 
(small symbols). Displayed are the dependence
on the outer- and inner electrolyte concentrations (Panels {\bf a} 
and {\bf c}), 
on the noise amplitude ({\bf b}), and on
the radius (width) of the system ({\bf d}). The values of
the parameters kept fixed in a panel, and the 
experimental estimate of the scale factor $\bar c$ 
are discussed in the text. 
Statistical errors are shown for the experiments 
and for a single set of simulations.}
\label{F:Exp-rev-multi}
\end{figure*}
Before describing the experiments further, we turn to the
theory since it allows a more concise discussion of the results. 
Theories of Liesegang-type patterns combine the properties of a moving front (i.e. where and at what rate the reaction product, $A+B\to C$, appears) with the details of the precipitation (i.e. how the reaction product, $C$, turns into precipitate). While the front properties have been 
thoroughly studied and understood both 
theoretically \cite{GR1988,MatPack98} and experimentally \cite{Bazant1999,Tabeling2003}, the dynamics of 
precipitation is more debated \cite{MullerRoss2003,MatPack98}. 
The competing pre- and post-nucleation views can be combined 
\cite{MatPack98,Heureux}, and we shall use a simple version 
\cite{ModelB1999} based on the Cahn-Hilliard equation with
noise added \cite{CahnHill1958,Halp-Hoh1977,Volfi}. 
This equation features spinodal-decomposition-type fast dynamics, 
as well as slower, nucleation-and-growth processes \cite{Gunton1983}. 
Driving it with a reaction zone gives us a flexible model 
with a variety of pattern-formation regimes. 

The reaction front appears due to a strongly inhomogeneous initial distribution of the reagents $A$ and 
$B$. The reaction takes place in a gel (occupying the half space $x>0$) and, initially, the inner electrolyte $B$ is homogeneously distributed [$b(x>0,y,t=0)=b_0$]. The outer electrolyte $A$ of much higher  
concentration [$a(x<0,y,t=0)=a_0$ with $a_0\gg b_0$] is 
brought into contact with the gel at $t=0$. Assuming a second-order irreversible reaction $A+B\to C$, the front invading the gel can be described by the equations
\begin{eqnarray}
\partial_t a&=&D_A\Delta a-kab
\label{e-1}\\
\partial_t b&=&D_B\Delta b-kab
\label{e-2}
\end{eqnarray} 
where both the reaction rate $k$ and the diffusion coefficients, 
which are assumed to be equal ($D_A=D_B=D$), are set to 1 
by an appropriate choice of the time- and length-scales 
\cite{epaps}. The front, specified in terms of 
the rate of production of 
$C$s $(kab)$, is narrow and moves into the gel diffusively
(the position is given by $x_f=\sqrt{2D_ft}$ where $D_f$  
is a function of $D$ and $b_0/a_0$).
The front leaves behind a constant concentration of $C$s $(c_0)$, 
where $c_0$ depends on $D$ and $b_0/a_0$, and it is practically 
independent of $k$. Provided the system with $c_0$ is unstable 
or metastable, a phase separation of $C$s into regions of 
high $(c_h)$ and low concentrations $(c_\ell)$ takes place. 
This process is described by the Cahn-Hilliard equation with source 
$(kab)$ and noise $(\eta_c)$ terms added
\begin{equation}
\partial_t m=-\lambda\Delta (m-m^3+\sigma \Delta m)+kab+\eta_c
\label{CHsc-eq} \, .
\end{equation}
Here $m$ is the concentration of $C$s shifted by $\bar c=(c_h+c_\ell)/2$ 
and scaled by $\hat c =(c_h-c_\ell)/2$, so that $m=(c-\bar c)/\hat c$ is 1
for $c=c_h$ and $m=-1$ for $c=c_\ell$.
The parameters $\lambda$ and $\sigma$ are the rescaled kinetic 
coefficient and surface tension, respectively \cite{ModelB1999,Halp-Hoh1977,Volfi}. 
Their ratio $\tau_u\approx\sigma/\lambda$ defines a characteristic timescale 
of the growth of unstable modes in precipitation. 
Comparing $\tau_u$ with the time the front passes through a region 
determines whether slow nucleation-and-growth or fast  
spinodal decomposition dominates the pattern formation.

Adding noise ($\eta_c$) is essential since 
the formation of helices is a symmetry-breaking process. 
Furthermore, the noise 
widens the available regions of the meta- and unstable states 
(see earlier morphological phase diagrams of 
Liesegang patterns \cite{Chopard}). 
Noiseless Cahn-Hilliard type dynamics where the 
front moves with {\em fixed velocity} have been much studied \cite{Hantz,Foard-Wagner}. In these cases, however, noise was 
present in the initial state, and complex morphologies
resulted from complex initial conditions or 
from complex motion of the reaction front.
Our model without the noise reproduces the properties 
of the regular Liesegang patterns \cite{ModelB1999,LieseRZ1999}. 
Inclusion of the bulk noise allows us to 
demonstrate the existence of helices, and understand the experimental
trends in their emergence.

From a theoretical point of view, the {\it tube-in-tube} experiments
are the easiest to describe. We can cut and open the cylinder as shown 
in Fig.\ref{F:helix-plain} and treat the thin layer as a
two-dimensional strip of width $L_y=2\pi R$, and length equal to the 
tube length $L_x$ \cite{3dsim}. 
Accordingly, eqs.(\ref{e-1}-\ref{CHsc-eq}) are
solved in a rectangle of size $L_x \times L_y$ with periodic boundary 
conditions in the $y$ direction and no-flux boundary conditions 
at the lower edge $(x=L_x,y)$. At the upper edge 
(the initial location of the front $(x=0,y)$, we use a 
slightly idealized boundary: the concentration of the 
outer electrolyte is kept at a constant value
$a(x=0,y,t)=a_0/\bar c$ while no-flux condition is adopted
for $B$ and $C$. The initial conditions reflect the 
experimental setup: 
$b(x>0,y,t=0)=b_0/\bar c$, $a(x>0,y,t=0)=0$, and $c(x,y,t=0)=0$. 
The discretized noise term $\eta_c$ is 
implemented by moving $C$s between neighboring sites at a rate
$\eta_c=\sqrt{c}r$ where $r$ is uniformly distributed in 
the interval $[-\eta,\eta]$. In the following, 
$\eta$ is called the amplitude of the noise. 

Our simulations indicate that both the 
Liesegang bands and the helices emerge in a wide
range of the parameters. There are, of course, some constraints,
e.g., $\eta$ must be sufficiently small for the phase separation 
to take place. 
Examples of simulations are shown in Fig.\ref{F:helix-plain} 
(rightmost two columns), where a Liesegang pattern 
and a helix are displayed \cite{epaps}.
A general feature of the simulations is that the chirality 
of the helices is random within the statistical error of 
100 independent simulations. This is in agreement with the experiments where, out of 96 helicoids/helices, the ratio of 
left- and right-handed patterns is 50/46. We consider this as experimental evidence that the macroscopic symmetry breaking is not driven 
by microscopic objects of given chirality.

To characterize the emergence of the helices quantitatively,
we collected data by 
varying $a_0$, $b_0$, $\eta$ and $L_y$, and determined
$P_{\rm H}$ from the outcome of 100
simulations with distinct random number sequences
for $\eta_c$. Since the kinetic coefficients $\lambda$ 
and $\sigma$ cannot be controlled externally, we kept 
them fixed ($\lambda =0.2$, $\sigma =0.8$) throughout 
the simulations.

First, we varied $a_0/\bar c$ and $\eta$ 
while keeping $b_0/\bar c =1$ and $L_y=64$ fixed. 
Figs.\ref{F:Exp-rev-multi}a,b shows 
that $P_{\rm H}$ is remarkably large, it increases
with $a_0$ and reaches $P_{\rm H}\sim 0.4-0.6$ for large $a_0/\bar c$.
Similar trend is also seen in the experiments. 
Since $a_0/b_0 $ determines the front motion, with 
larger $a_0$ corresponding to faster diffusion, an important  
conclusion from Fig.\ref{F:Exp-rev-multi}a is that 
fast motion of the front facilitates the emergence of helices.

Fig.\ref{F:Exp-rev-multi}a,b also show that no helices form 
even for larger $a_0/\bar c$ if the noise is too small. 
Increasing the noise first increases $P_{\rm H}$, then 
$P_{\rm H}$ saturates in the region $0.05<\eta <0.09$ 
and, finally, $P_{\rm H}\to 0$ due to the absence of phase separation 
above $\eta \approx 0.09$. 
Comparing Fig.\ref{F:Exp-rev-multi}b with experiments
is difficult since the link between 
$\eta$ and $T$ is through complex
changes in diffusion, reaction rates etc. 
Our experiments indicate that $P_{\rm H}$ increases with $T$. 
This is in agreement with Fig.\ref{F:Exp-rev-multi}b provided  
$\eta\sim T$ and the experimental $T$ corresponds to small values of 
$\eta$.

We also varied $b_0/\bar c$ and $\eta$  while fixing 
$a_0/{\bar c}=80$ and $L_y=64$ (Fig.\ref{F:Exp-rev-multi}c). 
The probability $P_{\rm H}$ was found to be maximal in the middle of the
spinodal region ($b_0/\bar c\approx 0.9-1.1$) where isotropic 
precipitation structures develop through fast-growing,
linearly unstable modes.
Comparing the simulations (Fig.\ref{F:Exp-rev-multi}c) with 
experiments is difficult since neither 
$\eta$ nor the experimental concentration 
scale, $\bar c = (c_h+c_\ell)/2\approx c_h/2$, are known. 
We estimated $c_h$ by assuming that all the precipitate 
was in the helices and all 
the $B$s reacted and turned into $C$s. This estimate 
left an apparent shift between the experimental and simulation 
points (Fig.\ref{F:Exp-rev-multi}c). 
The shift may well be the consequence 
of overestimating $c_h$ (e.g., not all the $B$s reacted, 
or the bands are wider than their optical width).

The effect of increasing width ($L_y$) is displayed in 
Fig.\ref{F:Exp-rev-multi}d. The experimental parameters are 
described in Fig.\ref{F:helix-exp}, while in simulations, we used 
$a_0/\bar c =15$, $b_0/\bar c =1$, and
$\eta =0.04$. The experimentally observed
lower threshold for the emergence of helices is clearly present 
($L_{yc}\approx 32$), 
and one can also recognize the trend that $P_{\rm H}$
first increases with $L_y$ and then decreases 
for large $L_y$. As in experiments, $P_{\rm H}$
decreases due to the proliferation of more complex structures.
Complexity builds up for large $L_y$ since more 
long-wavelength transverse modes (modes in the $y$-direction) 
can fit into the system. They are unstable modes of the 
Cahn-Hilliard dynamics facilitating the formation of 
more intricate patterns.

\begin{figure}[htb!]
 \includegraphics[width=\columnwidth]{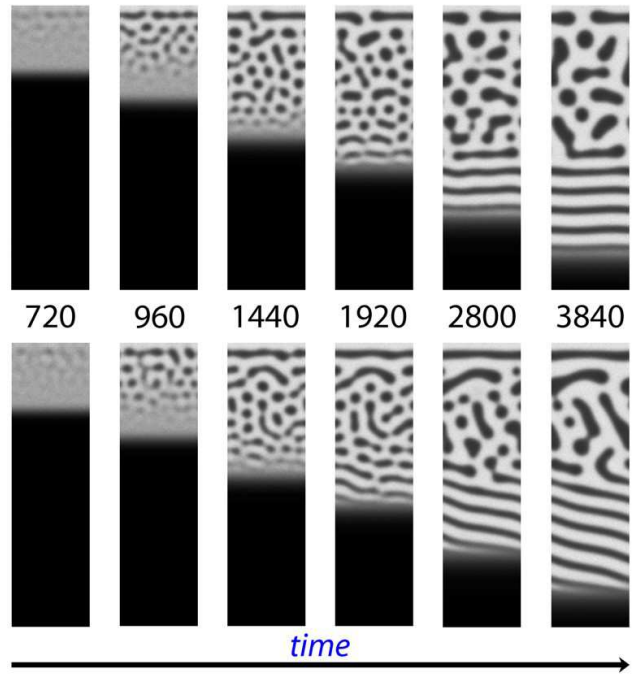}
\caption{Time evolution of the precipitate (white regions)  
for parameters 
$a_0/\bar c=60$, $b_0/\bar c=1$, $\eta =0.02$, and $L_y=64$. 
\label{F:helix-formation}}
\end{figure}

The common trends found in experiments and simulations suggest 
that our model contains the right ingredients, and we can develop 
a physical picture of helix formation by observing the simulations. 
Fig.\ref{F:helix-formation} displays two examples of time evolutions 
with parameters set to have roughly equal probabilities for bands 
and helices. There are many ways of choosing such parameters but 
the characteristic features of the dynamics are always 
the same. Essential among them is that, initially, 
the reaction front moves fast enough to produce a domain where the 
system is unstable and roughly homogeneous (fuzzy regions in the 
$t=720-960$ plates in Fig.\ref{F:helix-formation}). 
The homogeneity makes possible the 
generation of isotropic patterns which compete with the anisotropic 
influence of the front favoring band-formation perpendicular to 
the front motion ($t=1440-1920$ plates in Fig.\ref{F:helix-formation}). 
The outcome of 
this competition determines whether Liesegang bands, single 
helix, double helix or more complicated patterns emerge. 

One can quantify the above picture by noting that homogeneous patterns can form only if the front moves a distance of the order $L_y$ in a time, $\tau_f=L_y^2/2D_f$ that is smaller than the time, $\tau_u$, required for the precipitation to develop. To estimate $\tau_u$, we calculate the growth rate, $\omega_{k^*}=\lambda/4\sigma\approx 1/\tau_u$, of the fastest growing mode of wave-number $k^*=1/\sqrt{2\sigma}$ using linearized Cahn-Hilliard dynamics for a quench to the middle of the miscibility gap [$m(0)\approx 0$]. Then, assuming that the homogeneous structure emerges from the noise, we have $\sqrt{\eta}\exp{(\omega_{k^*}\tau_u)}\approx m(\tau_u)\approx 1$, and the inequality $\tau_f<\tau_u$ yields an upper limit for the width of a system $L_y^2< 4D_f\sigma |\ln{\eta}|/\lambda$ where helix can form. A lower limit can also be found since the characteristic size of the domains $(L^*\approx 2\pi/k^*=2\pi\sqrt{2\sigma})$ formed by the fastest growing modes must be smaller that the width of the system ($L^*<L_y$), otherwise no structure forms in the $y$ direction. The combination of the two inequalities 
\begin{equation} 8\pi^2\sigma < L_y^2 < 4D_f\sigma |\ln{\eta}|/\lambda \label{inequs} \end{equation} 
reflects some of the trends observed in the experiments and simulations. Namely, the formation of helices are facilitated by a fast moving front, i.e. by $D_f$ being large which, in turn, requires $a_0$ to be large and, furthermore, there is a minimal width below which no helices form.

Since the width ($L_y$) is bounded from both sides, it may happen that no helices can form. When searching for helices one should, in general, use a fast front (e.g., by selecting large $a_0$) and create an unstable state behind the front by placing the system deep in the miscibility gap (e.g., by experimenting with $b_0$). Finding the 
right temperatures is also important but it is a rather 
complex problem left for future studies.

Acknowledgments. 
This work was founded by the Hungarian Academy of 
Sciences through OTKA Grants No. K68109, NK100296, and 
K104990. IL was also supported by a Magyary Postdoctoral 
Fellowship. FM acknowledges
partial support by NSF through Grant No. DEB-0918413.

\onecolumngrid

\newpage

\vspace{0.3truecm}
{\section
{Supplementary Information}}
\vspace{0.3truecm}
\twocolumngrid
\appendix

\subsection{A. Detailed model description}
\label{Intro}

Liesegang patterns are formed in the wake of moving 
reaction-diffusion fronts. 
The reaction front emerges due to the inhomogeneous initial 
distribution of the reagents $A$ and $B$. Namely, the reaction takes 
place in a gel occupying the half space $x>0$ where, initially, 
the inner electrolyte $B$ is distributed homogeneously 
[$b(x,y,t=0)=b_0\theta(x)$].
The outer electrolyte $A$ of much higher initial concentration 
[$a(x,y,t=0)=a_0\theta(-x)$ with $a_0\gg b_0$] is 
brought into contact with the gel at $t=0$. 
Assuming a second-order, irreversible reaction 
$A+B\to C$, the front invading the gel is described by 
the equations
\begin{eqnarray}
\partial_t a&=&D_A\Delta a-kab
\label{e-11}\\
\partial_t b&=&D_B\Delta b-kab
\label{e-21}
\end{eqnarray} 
where $k$ is the reaction rate and, for simplicity, the
diffusion constants of the reagents are assumed to be equal
($D_A=D_B=D$). This assumption will be used throughout the 
two-dimensional simulations.

The front, characterized by the spatio-temporal 
properties of $kab$ (rate of production of $C$s),
has been studied in detail \cite{GR1988,MatPack98}. 
It is narrow and moves diffusively (its position
is given by $x_f=\sqrt{2D_ft}$ where $D_f$ can be expressed
through $D$ and $b_0/a_0$). It leaves behind a constant
concentration, $c=c_0$, of the reaction product $C$, where the
parameter $c_0$ is determined by $D$ and $b_0/a_0$, and 
is practically independent of $k$.

Assuming that no intermediate complexes are formed, the next stage
of the formation of the precipitation pattern is the separation of
the reaction product, $C$, into high- and low-concentration phases.
At a coarse grained level, the phase separation can be described
by the generalized Cahn-Hilliard 
equation \cite{CahnHill1958,Halp-Hoh1977,ModelB1999,Volfi}
\begin{equation}
\partial_t c=\lambda_0\Delta(\delta f/\delta c)+kab+\eta_{c0}
\label{CH-eq} \, .
\end{equation} 
Here $\lambda_0$ is a kinetic coefficient, $f$ is the free energy 
driving the phase separation, $kab$ describes the creation 
of $C$ particles by the front, and $\eta_{c0}$ represents 
noise effects (thermal fluctuations, inhomogeneities 
in the gel, etc.) which conserve the total number of $C$ particles.

In order to describe the phase separation, the free energy 
$f(c)$ should have two minima corresponding to the low- ($c_\ell$) 
and high ($c_h$) concentrations of $C$s 
in homogeneous equilibrium states. It
should also have a surface tension term preventing the 
formation of singularities at interfaces. As a convenient
form with minimal number of parameters,  
one can take $f$ as a Landau-Ginzburg free energy which is
symmetric about $\bar c =(c_h+c_\ell)/2$ 
\begin{equation}  
f(c)=-\frac{\varepsilon}{2}(c-{\bar c})^2+
\frac{\gamma}{4}(c-{\bar c})^4+\frac{\sigma_0}{2}(\nabla c)^2
\label{LGform} \, .
\end{equation} 
where $\varepsilon$, $\gamma$, and $\sigma_0$ are phenomenological 
parameters, and the minima of $f(c)$ are fixed at 
$c_h$ and $c_\ell$ by setting 
$\sqrt{\varepsilon/\gamma}=(c_h-c_\ell)/2\approx c_h/2$ where 
we use the fact that $c_h\gg c_\ell$ i.e. the gaps between the 
bands have very low steady-state concentration of $C$s
in the usual Liesegand experiments.

Measuring concentration, time, and length in units of
\begin{equation}
\hat c =\frac{c_h-c_\ell}{2}\approx \frac{c_h}{2} \quad , \quad \tau=\frac{1}{k\hat c}
\quad , \quad l =\sqrt{\frac{D}{k\hat c}}
\label{scales}
\end{equation} 
and, furthermore, making a shift in the concentration of $C$s
\begin{equation}
m=\frac{c-(c_h+c_\ell)/2}{(c_h-c_\ell)/2}\, \approx \, 
\frac{c}{\hat c} -1
\label{scaled-c} \, 
\end{equation} 
one obtains a simple set of equations 
\begin{eqnarray}
\partial_t a&=&\Delta a-ab
\label{sc-1}\\
\partial_t b&=&\Delta b-ab
\label{sc-2}\\
\partial_t m&=&-\lambda\Delta (m-m^3+\sigma \Delta m)+ab+\eta_c
\label{CHsc-eq-1} \, ,
\end{eqnarray}  
where $\lambda=\lambda_0\varepsilon /D$, 
$\sigma = \sigma_0k\hat c/D\varepsilon$, $\eta_c=\eta_{c0}/k{\hat c}^2$
are the rescaled kinetic coefficient, surface tension, 
and conserved noise, respectively. 

A few comments are in order about the random aspects of the 
dynamics. First, we note that randomness is not added
to the reaction-diffusion equations (\ref{sc-1},\ref{sc-2}) since
the noise has been shown to be irrelevant in determining 
the properties of the 
$A+B\to C$ type reaction fronts in the physically relevant 
dimensions \cite{cornell}. Second, we recall that 
the noise term $\eta_c$ in the generalized Cahn-Hilliard 
equation (Model B \cite{Halp-Hoh1977}) describes the local 
concentration-fluctuations resulting from
diffusive random motion of $C$s. This noise conserves 
the total number of particles and is expected to be proportional
to $\sqrt{c}$. 
Third, we point out that the amplitude of $\eta_c$ in 
near-equilibrium dynamics is proportional to the temperature 
and it is related to the kinetic coefficient $\lambda$. However, 
we have here a far from equilibrium situation, and
no fluctuation-dissipation relation connects these parameters.
Accordingly, we shall treat $\lambda$ and the amplitude $\eta$ 
of $\eta_c$ as independent parameters.  Of course, one expects that 
temperature is related to the noise and, in general, 
$\eta$ increases with temperature. In the simulations, 
the discretized noise term [$\eta_c$ in \eqref{CHsc-eq-1}] was implemented 
by moving $C$s between neighboring sites at a rate  
proportional to $\eta_c=\sqrt{c} r$ where $r$ is a uniformly
distributed random number 
from the interval $[-\eta,\eta]$, and $\eta$ is the parameter which is
called the amplitude of the noise.

The above formulation is remarkable 
in that the number of parameters ($\lambda\, ,\sigma\, , \eta$) 
is small compared to that found in the usual 
models of Liesegang phenomena. Among the parameters, $\sigma$
does not appear to be important since it just determines the
width of the transition between the high- and low-concentration 
regions. The parameter $\lambda$, on the other hand, 
does play an essential role 
since it sets the timescale of the precipitation processes. Comparing
this timescale with the time the front passes through a region
determines whether the nucleation-and-growth or the unstable growth
(spinodal decomposition) modes dominate the phase separation dynamics. 
Finally, the noise $\eta_c$ is also important. First, because the 
formation of helices is a symmetry breaking process which does not 
happen without the presence of noise. Second, because it determines 
the phase diagram (the meta- and unstable regions) for the given 
parameter values of the system.

Unfortunately, the parameters $\lambda\, ,\sigma\, , \eta$ 
are not easily controlled in experiments. For example, when trying
to amplify $\eta$ by increasing the temperature, one immediately 
realizes that there are a number of parameters 
(diffusion coefficients, reaction rate, etc.) 
which have strong but largely unknown temperature-dependence with 
unpredictable combined effects. 

The parameters which are contollable in the 
experiments come from the initial preparation of the systems. 
They are the 
initial concentrations of the electrolytes 
($a_0\,,  b_0$) and, furthermore, it turns 
out that the radius $R$ of
the test tube also sets some constraints on the emergence of
helices. 

\bigbreak

\subsection{B. Additional information about the simulations}

The solution of the discretized equations (\ref{sc-1}-\ref{CHsc-eq-1})
with the above boundary and initial conditions were carried out
using a uniform grid with various combination of scaled 
$L_x$ and $L_y$ from the ranges of 
$32\le L_y/l\le 512$ and $512\le L_x/l \le 2048$. 
The equations were integrated
in time by the simple Euler method (fast and extensive search 
in the parameter space was feasible by using the parallel 
programming possibilities of video cards). For the results 
quoted and displayed in the paper, the grid
spacing and the time step were 1.0 and 0.02, respectively.

\subsection{C. Three-dimensional simulations}

Three-dimensional systems can also be studied by 
simulating the $d=3$ versions of equations 
(\ref{e-11}-\ref{CH-eq}). There are two changes compared to the $d=2$ case. 
First, in order to be in agreement with the experimental 
setup, the periodic boundary conditions are replaced by 
no-flux boundary conditions in the direction perpendicular 
to the direction of motion of the front. Second, we 
lift the restriction of $D_A=D_B$ in our search for 
helicoids, and this means that an additional parameter 
$\theta =D_A/D_B$ appears in the scaled equations.
 
\begin{figure}[htb]
  \includegraphics[width=1.2cm]{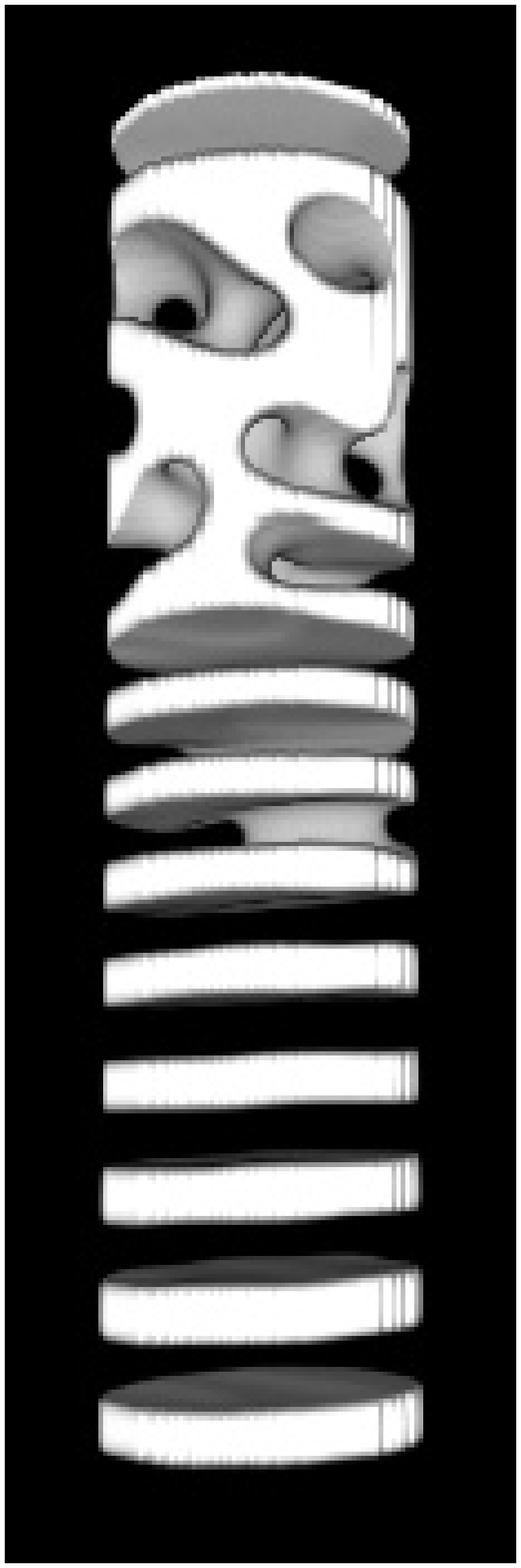}
  \includegraphics[width=1.23cm]{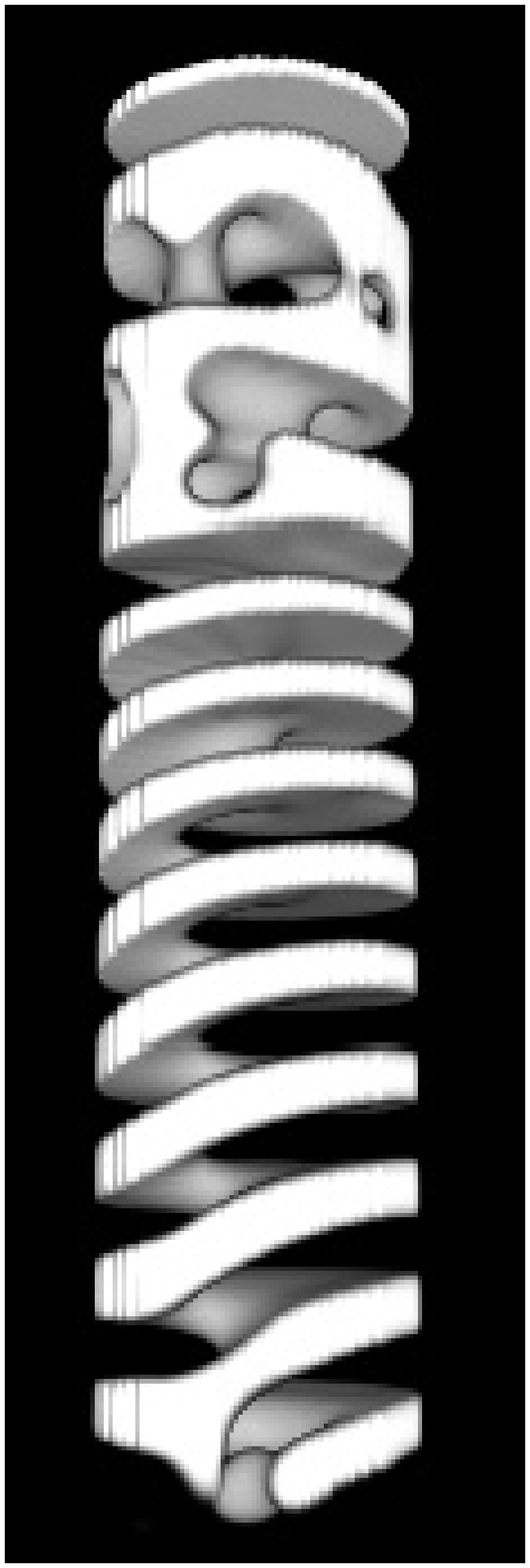}
\caption{Precipitation patterns (Liesegang on the left, helicoidal 
on the right) obtained in $d=3$ simulations using the 
following scaled parameters $a_0=100$, $b_0=1$, 
$\sigma =0.8$, $\lambda =0.2$, $\eta =0.02$, $\theta =1.2$, and $R=24$. 
The only difference in the simulations is the initial seed for the
random number generator. 
\label{F:helix-3d}}
\end{figure}

In the simulations,
we find that both the Liesegang and the helicoidal 
patterns observed in the
experiments can be qualitatively reproduced (see Fig.\ref{F:helix-3d}).
Unfortunately, the time-scale of the simulations compared to the $d=2$
ones is multiplied roughly by $2R$ where $R$ is the scaled radius of 
the tube (the grid
spacing used was 1.0). As a consequence, the computer power 
presently available to us is insufficient for obtaining good quality statistics for the probability of helicoid formation in $d=3$. 


\begin{thebibliography}{99}

\bibitem{Gao-Zincoxid} P. X. Gao, Y. Ding, W. J. Mai, W. L. Hughes, 
C. S. Lao, and Z. L. Wang, Science {\bf 309}, 1700 (2005).

\bibitem{Imai-chiral} H. Imai, and Y. Oaki, Angew. Chem. 
Int. Edit. {\bf 43}, 1363 (2004).

\bibitem{Su-doublehelix} D. S. Su, Angew. Chem. Int. Edit. {\bf 50}, 
4747 (2011).

\bibitem{Sci-1982-Muller} S. C. M\"uller, S. Kai, and J. Ross,
Science {\bf 216}, 635 (1982).

\bibitem{helical-ribbons} R. V. Suganthi, E. K. Girija, 
S. Narayana Kalkura, H. K. Varma, and A. Rajaram, 
J. Mater. Sci.- Mater. in Med. {\bf 20}, 131 (2009).

\bibitem{Giraldo2000} O. Giraldo, S. L. Brock, M. Marquez, S. L. Suib, 
H. Hillhouse, and M. Tsapatsis, Nature, {\bf 405}, 38 (2000).

\bibitem{heart-helix} P. Savadjiev, G. J. Strijkers, A. J. Bakermans,
E. Piuze, S. W. Zucker, and K Siddiqi, Proc. Nat. Acad. Sci. USA, 
{\bf 109}, 9248 (2012).

\bibitem{chiral-fibers} J. H. Jung, Y. Ono, K. Hanabusa, and S. Shinkai, 
J. Am. Chem. Soc., {\bf 122}, 5008 (2000).

\bibitem{chiral-blocks} Y. Oaki and H. Imai, J. Am. Chem. Soc., 
{\bf 126}, 9271 (2004). 

\bibitem{Grzy2005} B. A. Grzybowski, K. J. M. Bishop, C. J. Campbell, M. Fialkowski, and S. K. Smoukov, Soft Matter {\bf 1}, 114 (2005).

\bibitem{Carbontube} S. Y. Ju, J. Doll, I. Sharma, and F. Papadimitrakopoulos, Nat. Nanotechnol. {\bf 3}, 356 (2008).

\bibitem{Henisch} Henisch, H. K. {\it Crystals in gels and Liesegang rings}, Cambridge University Press, Cambridge  (1988).

\bibitem{MullerRoss2003} S. C. M\"uller and J. Ross, J. Phys. Chem. A {\bf 107}, 7997 (2003).

\bibitem{ModelB1999} T. Antal, M. Droz, J. Magnin, 
and Z. R\'acz, Phys. Rev. Lett. {\bf 83}, 2880 (1999).

\bibitem{Polezhaev1991} D. S. Chernavskii, A. A. Polezhaev, and S. C. M\"uller, Physica D {\bf 54}, 160 (1991).

\bibitem{Polezhaev1994} A. A. Polezhaev and S. C. M\"uller, Chaos, {\bf 4} 631 (1994).
  
\bibitem{GR1988} L. G\'alfi and Z. R\'acz, 
Phys. Rev. A {\bf 38,} 3151 (1988).

\bibitem{MatPack98} T. Antal, M.  Droz, J. Magnin, Z. R\'acz, and M. Zrinyi,
 J. Chem. Phys. {\bf 109,} 9479 (1998).

\bibitem{Bazant1999} C. L\'eger, F. Argoul, and M. Bazant, J. Argoul, 
J. Phys. Chem. B {\bf 103}, 5841 (1999).

\bibitem{Tabeling2003} C. N. Baroud, F. Okkels, L. M\'en\'etrier, 
P. Tabeling, Phys. Rev. E {\bf 67}, 060104 (2003).

\bibitem{Heureux} M. Chacron and I. L'Heureux, Phys. Lett. A {\bf 263}, 70 (1999).

\bibitem{CahnHill1958} J. W. Cahn and J. E. Hilliard, J. Chem. Phys. 
{\bf 28}, 258 (1958); J. W. Cahn, Acta Metall. {\bf 9}, 795 (1961).

\bibitem{Halp-Hoh1977} The Cahn-Hilliard equation with additive 
conserved noise is the much studied Model B of critical dynamics 
[P. C. Hohenberg 
and B. I. Halperin, Rev. Mod. Phys. {\bf 49}, 435 (1977)].

\bibitem{Volfi} A. Volford, I. Lagzi, F. Moln\'ar, and Z. R\'acz, 
Phys. Rev. E {\bf 80}, 055102(R), (2009).

\bibitem{Gunton1983} J. D. Gunton, M. San Miguel, and P. S. Sahni, in Phase Transitions and Critical Phenomena, edited by C. Domb and J. L. Lebowitz (Academic, London, 1983), Vol. 8.

\bibitem{epaps} See Supplementary Information for a detailed model 
description, and for examples of three-dimensional simulations.

\bibitem{Chopard} B. Chopard, P. Luthi, and M. Droz, 
Phys. Rev. Lett. {\bf 72}, 1384 (1994).

\bibitem{Hantz} P. Hantz and I. Bir\'o, 
Phys. Rev. Lett. {\bf 96}, 088305 (2006).

\bibitem{Foard-Wagner} E. M. Foard and A. J. Wagner, 
Phys. Rev E {\bf 85}, 011501 (2012).

\bibitem{LieseRZ1999}Z. R\'acz, Physica A {\bf 274}, 50 (1999).

\bibitem{3dsim} Computations for a 3D tube are 
also possible but exceedingly time consuming. An example of 
helicoid obtained in 3D is shown in \cite{epaps}. 

\bibitem{cornell} S. Cornell and M. Droz, 
Phys. Rev. Lett. {\bf 70}, 3824, (1993).

\end{thebibliography}
\end{document}